# Enhancing the capture efficiency and homogeneity of single-layer flow-through trapping microfluidic devices using oblique hydrodynamic streams

O. Mesdjian[1,2‡], N. Ruyssen[3‡], M.-C. Jullien[4], R. Allena[3], J. Fattaccioli[1,2*]

[1] PASTEUR, Département de Chimie, École Normale Supérieure, PSL University, Sorbonne Université, CNRS, 75005 Paris, France

[2] Institut Pierre-Gilles de Gennes pour la Microfluidique, 75005 Paris, France

[3] Institut de Biomécanique Humaine George Charpak, Arts et Métiers ParisTech, Paris, France

[4] Univ. Rennes 1, CNRS, IPR (Institut de Physique de Rennes) UMR 6251, F-35000 Rennes

‡Equal contribution

*Corresponding author e-mail: jacques.fattaccioli@ens.psl.eu

**ABSTRACT**

With the aim to parallelize and monitor biological or biochemical phenomena, trapping and immobilization of objects such as particles, droplets or cells in microfluidic devices has been an intense area of research and engineering so far. Either being passive or active, these microfluidic devices are usually composed of arrays of elementary traps with various levels of sophistication. For a given array, it is important to have an efficient and fast immobilization of the highest number of objects, while optimizing the spatial homogeneity of the trapping over the whole chip. For passive devices, this has been achieved with two-layers structures, making the fabrication process more complex.

In this work, we designed small microfluidic traps by single-layer direct laser writing into a photoresist, and we show that even in this simplest case, the orientation of the main flow of particles with respect to the traps have a drastic effect on the trapping efficiency and homogeneity. To better understand this phenomenon, we have considered two different flow geometries: parallel and oblique with respect to the traps array, and compared quantitatively the immobilization of particles with various sizes and densities. Using image analysis, we show that diagonal flows gives a spatial distribution of the trap loading that is more homogeneous over the whole chip as compared to the straight ones, and by performing FEM and trapping simulation, we propose a qualitative explanation of this phenomenon.

**KEYWORDS (4-6)**

Microfluidics, trapping, particles, simulation

## 1. INTRODUCTION

In order to monitor and analyze physico-chemical or biological processes, an important effort has been made in recent years to develop microfluidic devices (Nilsson et al. 2009; Narayanamurthy et al. 2017) for the control of the spatial positioning of small objects such as cells (Di Carlo et al. 2006a), bacteria (Eland et al. 2016), yeasts (Lee et al. 2008; Bell et al. 2014), droplets (Bai et al. 2010; Huebner et al. 2011; Pompano et al. 2011) or organoids (Murrow et al. 2017); at the level of a single object or in interaction with others(Dura and Voldman 2015). Immobilization or trapping can be achieved by different strategies : either by active methods such as the use of valves (Au et al. 2011; Zhou et al. 2016), droplet generation and arraying (Carreras and Wang 2017), etc. or by passive methods using hydrodynamic flows such as microwell arrays (Charnley et al. 2009). This article focuses on the last strategy.

Hydrodynamic trapping typically uses microfabricated mechanical barriers to create auxiliary flows that locally repel or immobilize target particles from the main flow. In the wide variety of devices that have been developed (Narayanamurthy et al. 2017), flow-through systems can be composed of auxiliary leakage channels regularly spaced perpendicular to the main flow of a serpentine channel (Tan and Takeuchi 2007; Jin et al. 2015; Zhou et al. 2016), or arrays of trapezoidal(Xu et al. 2013), half-circular (Di Carlo et al. 2006b) or U-shaped (Huebner et al. 2009) trapping pocket. Trapping arrays fabricated by single layer soft lithography (Wlodkowic et al. 2009; Chung et al. 2011; Yesilkoy et al. 2016; Zhou et al. 2016), could be improved in terms of efficiency, by making standing traps from double layer lithography (Di

Carlo et al. 2006a; Skelley et al. 2009), or in terms of selectivity, using reverse flow loading to immobilize multiple similar or different objects together (Skelley et al. 2009; Bai et al. 2010). In this work, we address the performances of single-layer trapping devices using a statistical approach. We show that the capture efficiency and homogeneity can be significantly improved by tilting the flow by an angle with respect to the longitudinal axis of the trapping array, rather than a using flow parallel to the array, as it is classically done. After a brief description of the fabrication process of microfluidic traps in straight and oblique devices, we quantify and compare the kinetics and homogeneity of trapping using model particles of comparable size to cells. Finally, using simple simulations and scaling arguments, we qualitatively explain the striking difference between the behavior of the straight and oblique configuration.

## 2. EXPERIMENTAL

### 2.1. Microfluidic trapping devices fabrication

Using standard soft lithography techniques (Xia and Whitesides 1998), we fabricate the SU-8 (SU-8 2015, Microchem) masters on silicon wafers by direct laser writing (Kloe Dilase 650, 375 nm) onto the photoresist, and development (SU-8 developper, Microchem). We then proceed to PDMS mixing (RTV 615, Momentive Performance Materials, 1:10 ratio), degassing, molding and thermal curing at 80 °C during two hours. We finally treat PDMS surfaces and glass coverslips (VWR, 26x76 mm) closing the channel with $O_2$ plasma (Femto science Cute, Operating conditions: 20 W, 50 kHz, 1 min) before bonding both parts of the chip together. The channels are filled with a solution of Poloxamer F-68 (0.1 % w/w) before injection of the particles or the droplets to avoid particle adhesion on the walls of the chamber.

### 2.2. Materials

Fluorescent polystyrene beads (DragonGreen, diameter of 5 µm) are purchased from Polyscience. Synperonic PE/F68 (Poloxamer 188, CAS 9003-11-6, $HO(C_2H_4O)_{79}$-$(C_3H_6O)_{28}$-$(C_2H_4O)_{79}H$) block-polymeric surfactant was kindly provided by Croda France SAS. Nile Red (CAS 7385-67-3), soybean oil (CAS 8001-22-7), and sodium alginate (CAS 9005-38-3) are purchased from Sigma-Aldrich. Ultrapure water (Millipore, 18.2 MΩ.cm$^{-1}$) is used for all experiments.

## 2.3. Emulsion droplets fabrication and staining.

(**5 µm diameter**) We first disperse 15 g of soybean oil in an aqueous phase containing 2.5 g of a surfactant (Poloxamer F-68, initial proportion of 30 %w/w) and 2.5 g of a thickening agent (sodium alginate, initial proportion of 4 %w/w) by manual stirring. This crude, polydisperse emulsion is further sheared and rendered quasi-monodisperse in a Couette cell apparatus under a controlled shear rate (5000 s$^{-1}$), following the method developed by Mason et al. (Mason and Bibette 1996). Before decantation, the emulsion is diluted in order to have a proportion of 1% w/w of Poloxamer F-68 and 5% w/w of oil. After one night of decantation, the oil phase is diluted with a solution of Poloxamer F-68 with an initial proportion of 1% w/w. After several decantation steps to remove very small droplets, the emulsion (final proportion of 50% w/w of oil) is stored at 12°C in a Peltier-cooled cabinet. (**14 µm emulsion sample**) The 14 µm mean diameter droplets are fabricated with a flow-focusing microfluidic device with a height of 5 µm and a width for the bifurcation channels of 8 µm. Soybean oil and a solution of Poloxamer F-68 (0.1 % w/w) with sodium alginate (0.1 % w/w) are injected with a pressure controller (pressures of the order of 300 mbar) for a flow rate of 1 µL per hour. (**Emulsion droplets staining**). After having prepared a 15 mg.mL$^{-1}$ Nile Red solution in DMSO, 1 µL of this mother solution is added in 100 µL of concentrated emulsion. Droplets are used after at least 1h of incubation at room temperature and rinsing of the external aqueous phase with the working buffer.

## 2.4. Microscopy.

Brightfield and fluorescent images particles are acquired on a Leica DMI8 microscope (Germany) connected to an Orca Flash 4.0 sCMOS camera (Hamamatsu Photonics, Japan). Epi-illumination is done with a LED light (PE-4000, CoolLED) and a GFP filter set (Excitation wavelength: 470 nm, Emission wavelength: 525 nm) for the fluorescent polystyrene beads, and a Cy3 filter set (Excitation wavelength: 545 nm, Emission wavelength : 605 nm) for the dyed fluorescent droplets. Time zero of the experiment is defined when the first particle is immobilized by a trap within the array. The number of particles per trap is measured by computing the ratio between the total fluorescence intensity within a region of interest corresponding to the trapping area for each trap and the average intensity of a single particle.

## 2.5. Layout design and image analysis.

Mask layouts are designed with WieWeb CleWin software. Image processing and analysis were done with Fiji/ImageJ (Schindelin et al. 2012). Data processing and analysis are performed with Mathworks Matlab software.

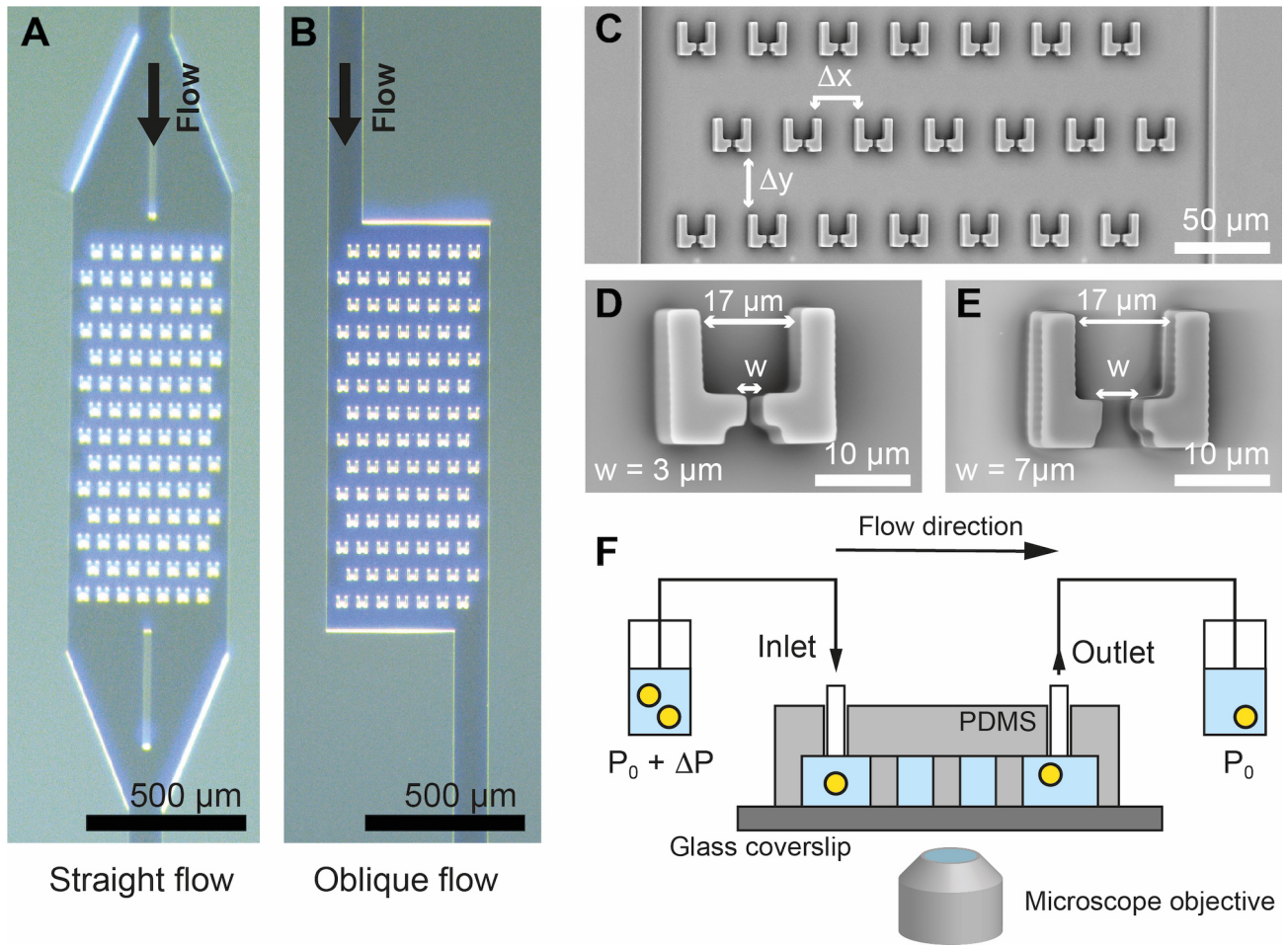

**Figure 1 :** Low-magnification pictures of the SU-8 masters for the case of the straight (A) and oblique (B) microfluidic chamber. The trapping part is composed of 7x14 staggered traps and has an overall dimension of 0.5x1.5 mm. Chamber height is equal to the trap height, $h = 14$ µm. Diverging parts upstream and downstream the trapping array are added in the straight design, to ensure that the streamlines are parallel to the longitudinal axis of the trapping array, and to avoid the collapse of the chamber ceiling. Flow direction is indicated by the arrows. (C) SEM mid-magnification view of the trapping array. Spacing between the traps is set to $\Delta x = 30$ µm laterally and $\Delta y = 50$ µm longitudinally. Individual traps have a rectangular chamber of 17 x 20 µm and a backside opening of (D) w = 3 µm or (E) w = 7 µm. The height of the chamber is set to 14 µm. (F) Schematic view of the experimental setup. Microfluidic chips and flowing/trapped particles are observed by brightfield and epifluorescence microscopy. Both the inlet and the outlet of the microchambers are pressurized using a computer-controlled pressure regulator.

## 3. RESULTS

## 3.1. Design of microfluidic trapping chambers.

Single-level microfluidic trapping flow-through chambers are manufactured by direct laser writing onto a 14 µm thick SU-8 photoresist (**Figure 1A**), followed by PDMS casting and sealing with glass after plasma treatment. Traps are U-shaped and are staggered in a 7 x 14 elements array, as shown in **Figure 1B**, with a lateral distance $\Delta x$ = 30 µm between the traps and an interline distance $\Delta y$ = 50 µm. Traps inter-distance within the arrays have been chosen in accordance to the design rules reported in the work of Skelley *et al.* (Skelley et al. 2009). Each individual trap has a 20 x 17 µm rectangular trapping cup, and a small opening at the back (see **Figure 1C**) that allows liquid to flow through the cup and ultimately trap objects (Huebner et al. 2009). The height of the microfluidic chamber is set to 14 µm by the height of SU-8 photoresist layer. Backside opening dimensions are chosen in accordance with the dimensions of the particles to be trapped: w= 3 µm for 5 µm particles or droplets, and w= 7µm for 14 µm large droplets. The trap arrays are inserted into two different microfluidic chip designs: a straight chamber (**Figure 1A**) and an oblique chamber (**Figure 1B**). Such configurations impose a large-scale flow orientation to the trap array, respectively parallel and diagonal to the longest axis of the device. In the case of the oblique chamber, the angle between the inlet and the outlet is set to 20° with respect to the long axis of the chamber.

## 3.2. Experimental setup and seeding particles.

By connecting the microfluidic devices to a pressure controller, a fixed pressure drop $\Delta P$ between the inlet and the outlet of the chamber is set. For a Newtonian fluid, this corresponds ensuring a constant flow during the experiment. Particle displacements and trapping are observed by video-microscopy.

To study the trapping efficiency, homogeneity and kinetics of our devices, two types of particles are used: 5 µm commercial polystyrene microbeads and homemade soybean oil-in-water emulsion droplets having an average size of 5 ± 1.7 µm and 14.4 ± 1.0 µm (similar to the one used in (Molino et al. 2016)). Note that for the bigger droplets, their diameter is of the same size as the channel height. Therefore, they cannot have the same speed as the carrier fluid since dissipation is then localized in the menisci. Studies for pancake droplets show that the velocity of the drops is then a fraction of the velocity of the carrier phase (Reichert et al. 2019). This field is still very active and to our knowledge there is no consensus on the prediction of the velocity of drops when they are in contact with walls. Nevertheless, we believe that this quantitative measure influences filling kinetics marginally and is neglected when quantifying trapping efficiency.

We chose to work at a particle concentration in the $10^6$ mL$^{-1}$ range, according to the values reported in the literature for the mammalian cell trapping experiments (Narayanamurthy et al. 2017), and corresponding to the concentrations commonly achievable with standard cell culture protocols without further enrichment or pre-concentration steps. This corresponds to a diluted regime with in average less than 10 particles flowing in the chamber at the same time (see **Supplementary Information** for the precise estimate). Polystyrene particles are purchased with a fluorescent dye conjugation and for soybean oil emulsion, a tiny amount of Nile Red, a lipophilic dye, is dissolved in the hydrophobic core of the droplets (Molino et al. 2016). In order to study the trapping efficiency with a statistical approach, we chose the dimensions of the traps so that at least one object can be captured. The quantification of the number of particles immobilized in the trap is performed by image processing, as detailed in the **Experimental** section.

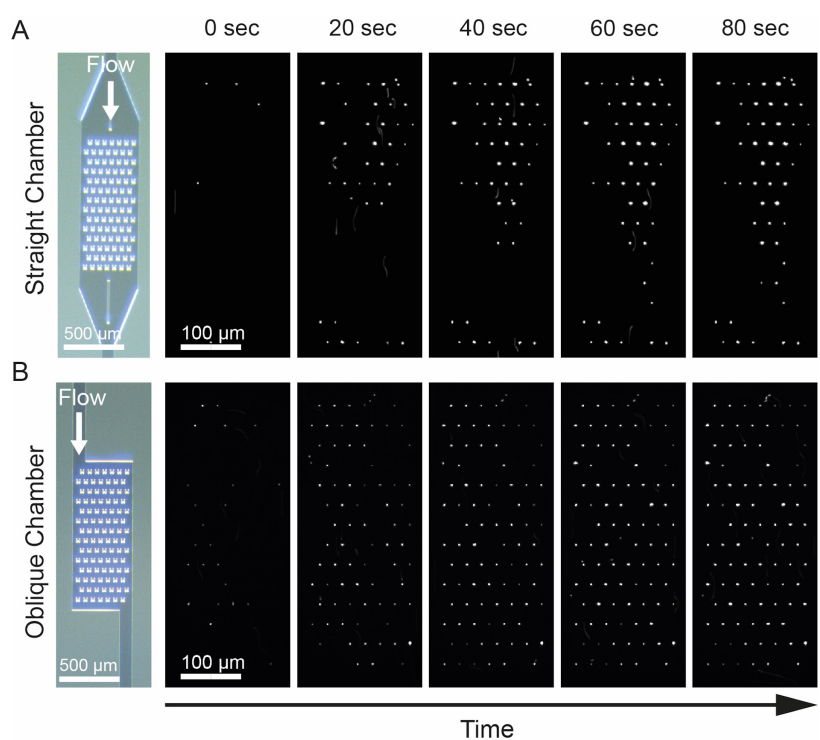

**Figure 2 :** Epifluorescence time-lapse imaging of chamber loading by 5 µm fluorescent polystyrene particles for (A) straight and (B) oblique designs. Individual traps have a backside opening width w = 3 µm. The flow rate is 0.5 µL.min$^{-1}$ and the particle concentration is $10^6$ mL$^{-1}$.

## 3.3. Particle trapping is more efficient in an oblique chamber.

In order to qualitatively evaluate the influence of trap orientation, typical particle capture experiments over time are shown in **Figure 2**, obtained by epifluorescence of a suspension of polystyrene particles. For both orientations (straight and oblique), the flow velocity, measured by analyzing the beads displacements in the chamber, is about 1 mm.s$^{-1}$ for a pressure drop of ΔP=50 mbar. At first observation, it is obvious that the filling of the trap in the oblique chamber is spatially more homogeneous than in the straight chamber. Moreover, the complete filling of the traps is done more quickly than in the case of the straight chamber. The trap filling rate, obtained by image processing, is defined as the number of traps containing one or more particles divided by the total number of traps, as a function of time. This quantification is performed for 5.0 and 14 µm oil droplets, and 5.0 µm polystyrene particles.

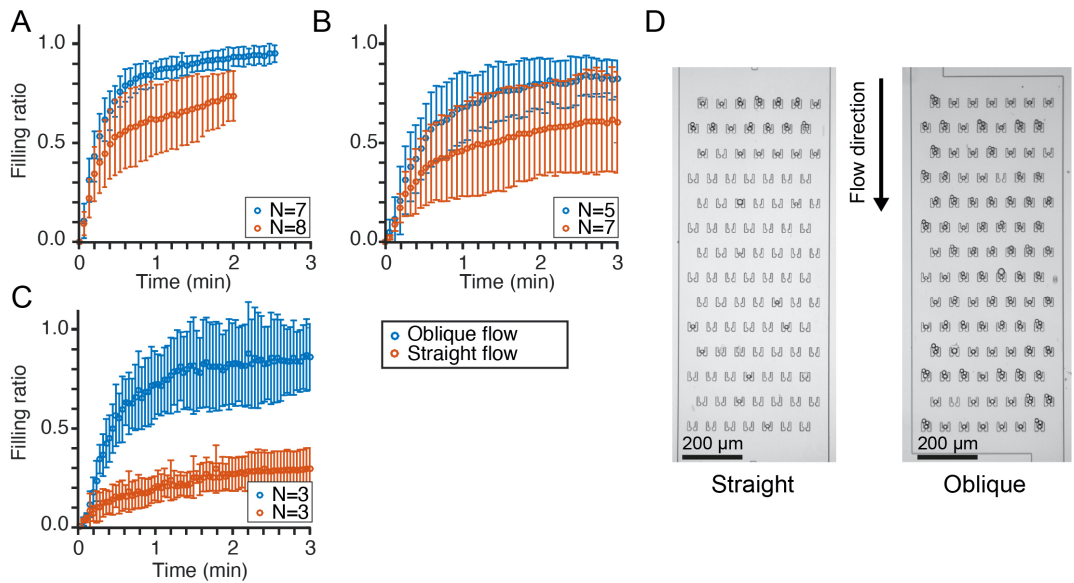

**Figure 3**: Comparison of the kinetics of filling ratio of the chambers for straight (red) and oblique (blue) chambers, for the case of (A) 5.0 µm polystyrene particles, (B) 5.0 µm quasi monodisperse emulsions droplets and (C) 14.4 µm monodisperse emulsion droplets. Particle concentration is set to 10$^6$ mL$^{-1}$ for the (A) condition and 3.10$^6$ mL$^{-1}$ for both (B) and (C) conditions. For (A) and (B), traps with a backside opening w = 3 µm are used, whereas traps with backside openings w = 7 µm are used in (C). The pressure drop ΔP is set to 50mbar. Each experimental curve is the average of N experiments, and the error bars correspond to the standard deviation of the experimental data. Representative microscopic image of a straight (D) and oblique (E) chamber taken after 3 min of injection of 14 µm large emulsion droplets. Injection direction is indicated with an arrow.

**Figure 3** shows that, regardless of particle type and size, the trap filling rate rapidly converges to a value close to 1 in about 1 min for the oblique design, meaning that all traps contain at least one particle at the end of the experiment. This complete filling is not achieved for the straight chamber which rather seem to converge to a finite value of filled traps. For 5 µm polystyrene particles (**Figure 3A**) and emulsion droplets (**Figure 3B**), this improved efficiency is accompanied by a smaller standard deviation of the filling rate kinetics in the case of the oblique chamber compared to the straight chamber. For large 14.4 µm oil droplets, the difference in efficiency is even more pronounced than for smaller particles, as shown in both the graphs in **Figure 3C** and the microscopic images recorded after 3 min of suspension injection. As a whole, for a given size of an object, it is difficult to conclude whether droplets are less trapped than PS beads. However, it is clear that the size of the objects plays an significant role in the process. Importantly, when droplets have a size similar to the one of the traps, the trapping efficiency is enhanced by a factor 3 using an oblique flow as compared to the straight one.

### 3.4. Efficiency is associated to a homogeneous filling

Since both polystyrene and emulsion particles are fluorescent, a numerical integration of the fluorescent intensity can be easily converted in an effective particle number by dividing the total intensity within a specific trap by the intensity value measured when the trap encloses a single particle. To better visualize locally the spatial homogeneity of the filling over the whole chamber, the trap array is converted, for the sake of representation, into a 7 x 14 pixel filling map array, where each pixel represents a single trap and is color-encoded with respect to the number of particles immobilized within it. **Figure 4** shows that for polystyrene particles, spatial distribution of particles is highly inhomogeneous with a straight chamber, traps positioned in the center being filled by a single particle, whereas traps positioned at the entrance and the exit of the trapping array can contain up to 6 particles.

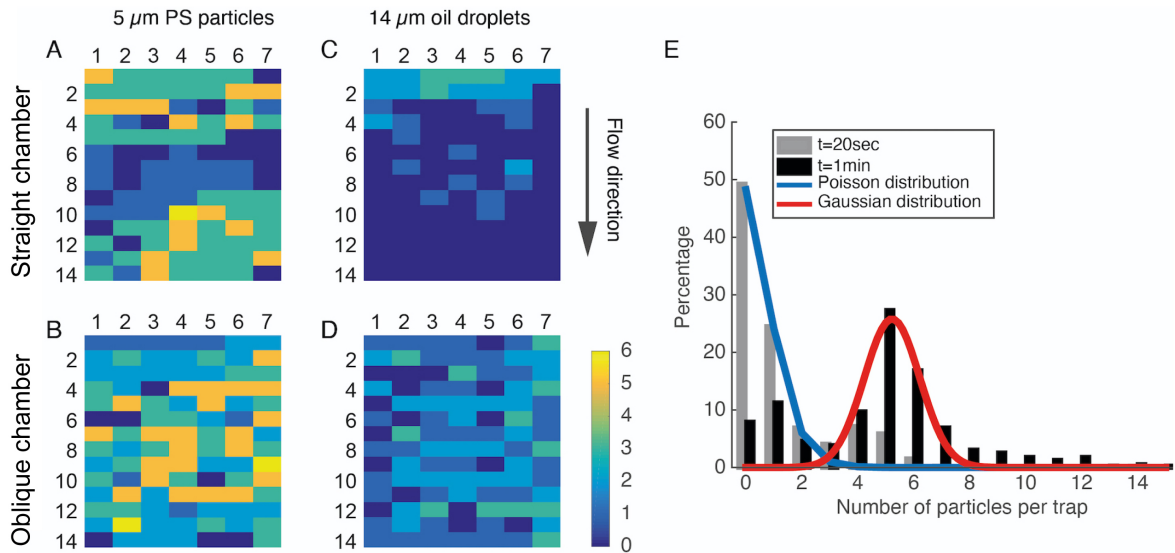

**Figure 4 :** Filling maps for straight (A, C) and oblique (B, D) chambers for the case of 5 μm polystyrene particles (A, B) and 14 μm monodisperse emulsion droplets (C, D) respectively. Each trap is represented by a colored cell in the arrays, which is color coded according to the number of particles it contains. Maps are plotted from experimental data recorded after 40s of injection. Particle concentration has been set to $10^6$ mL$^{-1}$ for (A, B) and $3.10^6$ mL$^{-1}$ for (C, D). Traps with a backside opening w = 3 μm were used for polystyrene particles, and traps with w = 7 μm for emulsion droplets. Injection pressure drop ΔP was set to 50mbar. (E) Comparison of the particle trapping distribution in a oblique chamber for 5 μm polystyrene beads, 20 s (blue) and 1 min (yellow) after injection. Particle concentration has been set to $3.10^6$ mL$^{-1}$. Traps with a backside opening w = 3 μm were used, with an injection pressure drop ΔP=50mbar. Histograms have been built from N=4 independent experiments. Data are fitted by a Poisson distribution ($\lambda$ = 0.5 ± 0.05 ) at short time scales, and by a Gaussian distribution (average of 5.2 ± 1.0 SD) at long time scales.

On the contrary the number of particles per trap in an oblique chamber ranges between 0 to 6 particles no matter its position within the array. In the case of 14.4 μm emulsion droplets, only the traps close to the entrance of the straight chamber are filled, with a maximum value of 3 droplets per trap, whereas for the oblique chamber, and similarly to the former case of PS particles, traps are spatially homogeneous, with a number of droplets ranging from 0 to 3 droplets over the whole chip, with a majority of traps enclosing one or more droplets. **Figure 4E** shows that at short timescales, in the case of PS particles flowing in the oblique devices, the particle trapping distribution follows a Poisson distribution with an average particle per trap $\lambda$=0.50 ± 0.05. At longer timescales, the particle trapping distribution follows a normal distribution with an average $\lambda$=5.2 ± 0.1 particles per trap. From these results, we can envision two strategies to trap single particles within the trapping array : (i) adapting the size of the traps to the dimensions of the particles so the traps cannot contain in average more than 1 particle, as in **Figure 4C** and **D,** or (ii) playing on the kinetics of the trapping statistics, and work in the Poisson regime of **Figure 4E.**

## 4. NUMERICAL SIMULATIONS

In order to explain the difference of spatial and temporal efficiency between the two chambers, we supplemented the experimental results with numerical simulations based on fluid-structure hydrodynamic interactions, in order to qualitatively identify the flow structure differences between the two configurations, and excluding additional physical mechanisms like gravity or short-range interactions between particles and walls.

The flows are simulated using COMSOL Multiphysics (5.3 version). In the following, we present the inputs that are used for the simulations, namely, the model, the channel and traps geometries, the carrier fluid and the objects properties, and finally the boundary conditions. The geometry is built from the nominal dimensions of the microfluidic chip. To avoid the heavy computations needed to have a complete 3D modelling, we chose to perform a less demanding 2D simulation since the width and length of the chamber are large compared to its height (Stone 2007). Despite the simulation being performed in 2 dimensions, we consider the third dimension using a Darcy approximation to account for the shear in the out-of-plane dimension.

Inside the meshed domain, we assume steady flow at low Reynolds number, which, for a Newtonian fluid reduces to the Stokes equation:

$$\mu \underline{\Delta}(\underline{V}) - \underline{\nabla}(P) + \underline{f_v} = \underline{0} \tag{1}$$

Where $\mu=10^{-3}$ Pa.s is the fluid dynamic viscosity, $\underline{\Delta}$ the first order Laplace operator, $\underline{V}$ the fluid velocity field, $\underline{\nabla}$ the first order gradient operator, $P$ the pressure field and $\underline{f_v}$ the force per unit of volume. Experimentally, we work with a particle concentration low enough to consider the solution as Newtonian and neglecting the particle-particle interaction. Calculations about this last hypothesis are exposed in the supplementary information section. We thus assume that the computed streamlines are not affected by the presence of particles. At the end, we focus on the streamlines passing through a trap and count their occurrence.

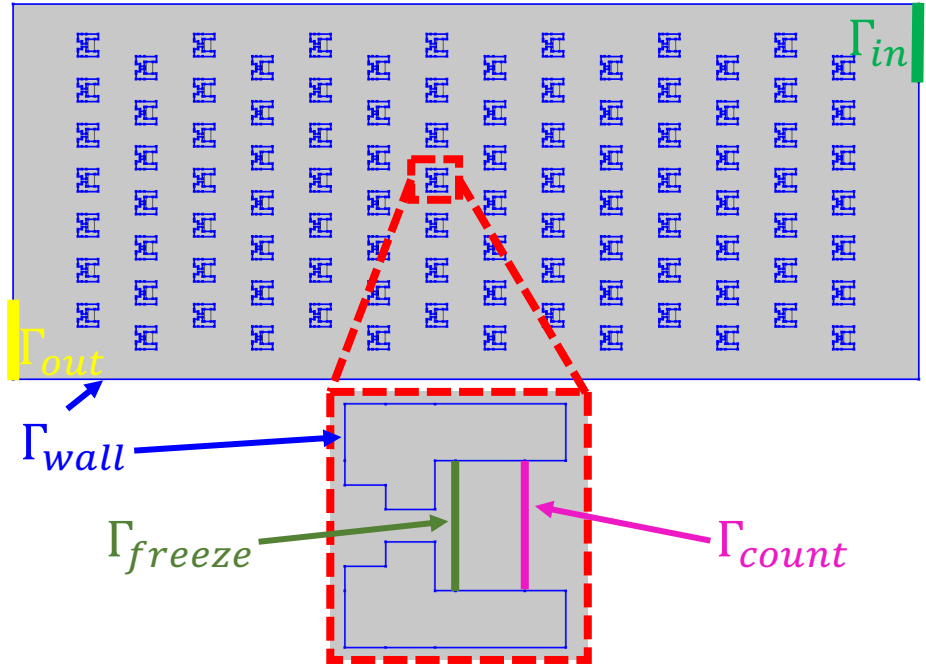

**Figure 5 :** Numerical model's geometry and its different boundaries. $\Gamma_{in}$ is the inlet boundary where a pressure of 100 Pa is considered, $\Gamma_{out}$ the outlet boundary where a pressure of 0 Pa is considered, $\Gamma_{wall}$ the walls boundaries with a no-slip condition applied, $\Gamma_{freeze}$ the boundary that stops the streamline passing through the trap mimicking a particle trapping the particles and $\Gamma_{count}$ the boundary that allows counting particles the streamlines entering the traps.

The flow is considered incompressible, the fluid mass conservation thus writes:

$$div(\underline{V}) = 0 \qquad (2)$$

where $div$ is the zero-order divergence operator.

To account for the shear in cross-section of the channel, we introduce the friction force per unit volume:

$$\underline{f_v} = \frac{-12\,\mu}{d^2}\,\underline{V} \qquad (3)$$

Where $d$ is the chip out-of-plane height ($d = 14\,\mu m$). Gravity does not contribute to the flow as the channel is placed horizontally (it simply generates a hydrostatic pressure in the out-of-plane direction).

The description of the geometry and the boundaries is shown in **Figure 5**. On the inlet boundary ($\Gamma_{in}$) a uniform pressure is imposed and equals to the experimental pressure drop ($\Delta P \approx 100\,Pa$) between the inlet and the outlet of the chamber. At the inlet, we consider a set of $N_s = 200$ streamlines with uniformly distributed intersection point with ($\Gamma_{in}$) along the whole length of ($\Gamma_{in}$). At the outlet boundary ($\Gamma_{out}$), where streamlines that do not cross the traps or immobilized are flowing out of the geometry, we set the pressure to zero. In addition, we consider a no-slip boundary condition on all the physical walls ($\Gamma_{wall}$) composing the microfluidics chamber. Since the traps are not physically fulfilled, streamlines are not interrupted at the backside opening of traps. To circumvent this issue, and to quantify the spatial distribution of streamlines crossing traps, we add two set of virtual boundaries for each trap of the geometry, as shown in **Figure 5** : (i) an inner boundary that counts the number of streamlines ($\Gamma_{count}$) entering the trap, and (ii) an inner boundary that store and display the intersection point position between a streamline and it ($\Gamma_{freeze}$).

The geometry is meshed by linear triangular elements, as shown in Erreur ! Source du renvoi introuvable.. To solve the equation set detailed above, the Stokes equation was spatially discretized with the Finite Element Method (FEM) and solved with a Parallel Direct Sparse Solver (PARDISO).

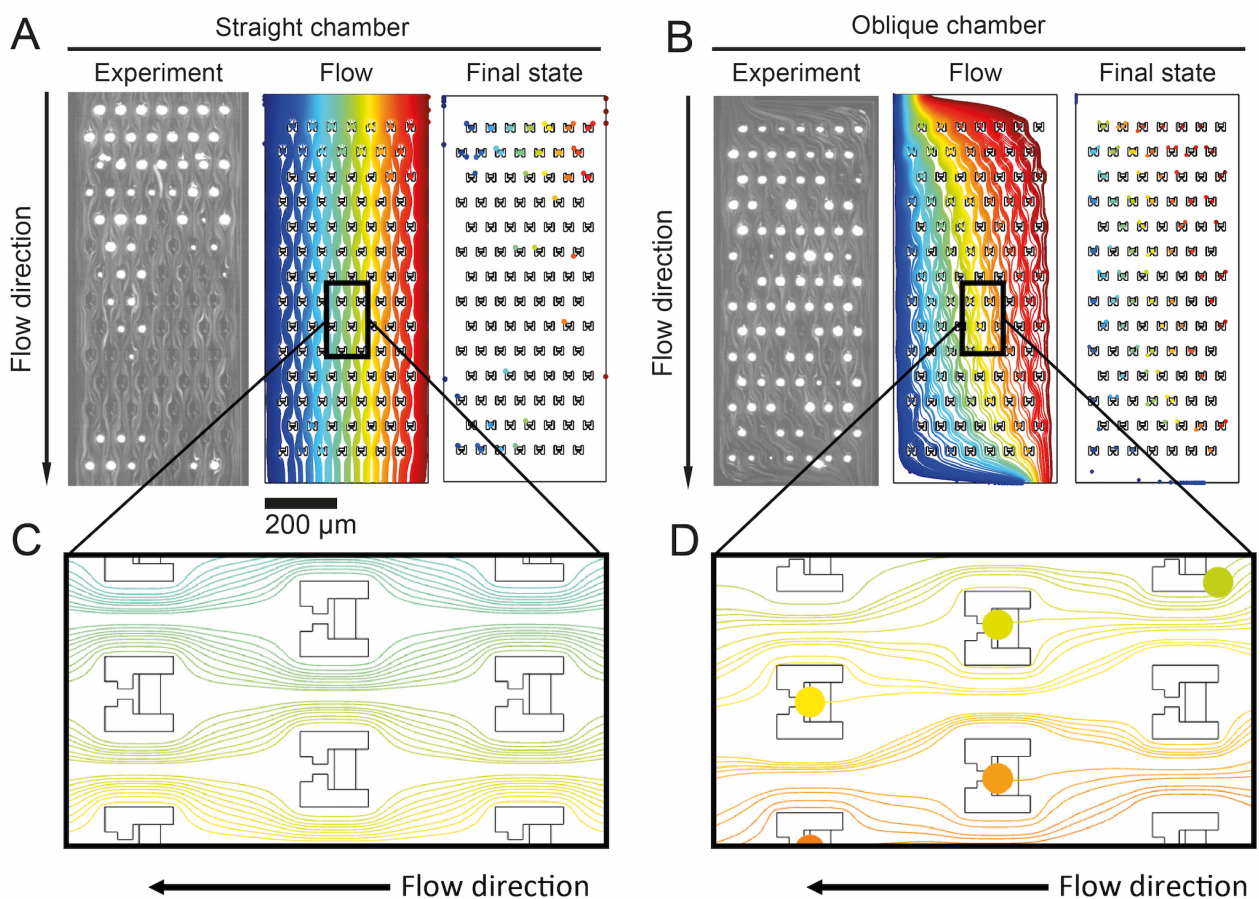

**Figure 6** : Comparison of trapping experiments with (A) straight and (B) oblique chambers, to FEM flow and trapping simulations. (C) and (D) are the enlarged 90° rotated view of respectively (A) and (B) in the central region of the chamber. Experimental pictures have been performed with PS particles, and the contrast has been adjusted to visualize the trajectories of the particles, which highlight the streamlines. On the simulations, colors correspond to intersection point positions between streamlines and the inlet boundary (dark blue: top left, dark red: top right). Trapping simulations have been performed with 200 streamlines. The streamlines are not physically interrupted by traps but when they cross a trap, the position of their intersection points with ($\Gamma_{freeze}$) is marked by a disc of the same diameter as the experimentally used particles. To facilitate readability, the portions of streamlines crossing traps after the trapping site are not displayed. When a streamline is physically stopped on ($\Gamma_{wall}$), its stop point is also marked by a disc of the same diameter as the experimentally used particles.

## 5. DISCUSSION

The experiments performed with the straight and oblique chambers show that in the latter case, the trap filling process is more rapid and spatially more homogeneous. The averaged velocity value in the chamber, measured by microscopy and in the mm.s$^{-1}$ range, is similar for both designs, and get reduced tenfold within the traps due to the small size of the backside opening, as shown by the line profile in
.

We show that a numerical model simulating fluid-structure interaction reproduces the phenomenological observations. This drives us to the conclusion that other mechanisms such as short-range interactions and sedimentation effects that may intervene are neglected.

In order to understand the role of the flow on the trapping efficiency with respect to the cavity geometry, we first consider the shape of the streamlines within the chambers (**Figure 6**).

As expected for a low Reynolds flow (Re ~ 10$^{-2}$ for a velocity in the mm.s$^{-1}$ range, an chamber height of 14 µm and a 30 µm distance between the traps), the computed streamlines are laminar, circumvent the obstacles, and agree with the experimental trajectories of fluorescent polystyrene particles (**Figure 6A**, **B**).

In addition, when looking at the spatial repartition of streamlines passing through traps obtained in the simulation and in the experiments at steady state, we observe the same loading pattern for both the straight and the oblique chamber (see **Figure 6**). More precisely, in the case of the straight chamber, we notice the appearance of an "empty" region where no particles are trapped in the middle of the chamber. The striking similarities between the results of the experiments and those of our model validate our approach based on hydrodynamics.

It is possible to understand how the flow is associated to the capture process by looking more closely at the shape of the streamlines. As depicted in **Figure 6A**, we show that the flow displays an axis of symmetry upstream/downstream in a transverse cross-section. However, by introducing an angle, the upstream/downstream symmetry is broken in the oblique geometry as depicted on **Figure 6B**. This symmetry breaking is at the origin of different capture efficiency as discussed below. Indeed, if a particle of fluid is located in the streamlines beam that is diverted from the trap upstream of the first trap, the stagnation points being located at angles 0 and $\pi$ of the trap, the probability that it will find itself in the capture streamlines beam of the next trap is very low, considering the symmetry of the streamlines. In order to allow the capture, it is therefore necessary to break the upstream/downstream streamline symmetry relative to the horizontal, which is precisely what the oblique chamber allows, as shown on the detailed views of the two configurations shown in **Figure 6C, D**.

To verify some assumptions made on the numerical model, several supplementary calculations were performed. First of all, the laminar regime hypothesis was verified using the Reynolds number value between traps: as its value is very small with respect to 1 we can confirm this hypothesis (for calculation details see the supplementary information section). Secondary, we estimated the average distance between particles considering a uniform particles distribution inside the chamber. As this average distance is greater than 10 times the particles diameter, we can confirm that the particle-particle flow interactions are negligible (for calculation details see the supplementary information section). We also investigated the effect of trapping on the streamlines shape: **Figure S 4** shows that two-dimensional simulations with an interrupted fluid flow inside the traps reveal that the effect of trapping is very weak on the streamlines outside the trap's region. Indeed, the streamlines have the same shape as in **Figure 6**. Moreover, even though trapping increases the hydrodynamic resistance of the microfluidic chamber, we notice that its effect on fluid velocity field is very weak (see **Figure S 3** and **Figure S 6 for comparison).** This is probably because the initial hydraulic resistance of traps is already very small (before trapping) with respect to the hydraulic resistance of the channels between traps. We also calculated the total flow rate going through the chamber and deduced the hydraulic resistance of the chamber when traps are fulfilled and when they are empty. As expected, the effect of trapping on the hydraulic resistance is small: 4,6% of variation for the straight chamber (with a variation of hydraulic resistance $\Delta R_{straight} = 8.10^{-11}$ Pa.s.m$^{-3}$) and 2,7% of variation for the oblique chamber ($\Delta R_{oblique} = 7.10^{-11}$ Pa.s.m$^{-3}$)

Three-dimensional (3D) simulation obtained with an extruded mesh (**Figure S 7**) confirms that the fluid flow is planar. To verify this, we simulated the fluid flow without Darcy's law and observed the shape of two layers of 100 streamlines crossing the inlet on uniformly distributed points. The upper layer of streamlines (red layer **Figure S 8**) is located at ¾ of the total height of the chamber, the lower layer of streamlines is located at ¼ of the total height of the chamber. We notice in the planar vertical view (**Figure S 9A1** and **B1**) that the lower layer (blue) overlaps with the upper one (red) excepted for 1% of the streamlines which is due to a mesh convergence issue. This result means that there is almost no planar deviation between pairs of streamlines. The planar horizontal view shows also that the vertical distance between the two layers of streamlines is also constant. these considerations suggest that a 2D approach is sufficient. Our model however presents some limitations: in the real device particle-wall interaction can occur. If a particle encounters a wall, its center of mass trajectory maybe deviated towards another streamline. Thus, the particle trajectory cannot be considered as identical to a streamline. To tackle this issue, an improvement would be to consider the particles dimensions in the fluid flow and the flow modification due to particles motion at each time step. Such a study should be perform using more advanced numerical methods like the Arbitrary Lagrangian Eulerian (ALE), or level set & phase field approach. However, the ALE method is known for its mesh distortion issues when particles come in contact to the walls, even with adaptative meshing tool. Level set & Phase field methods also becomes a problem when the particles size is very low with respect the entire fluid domain because of stability issues in addition to high computational cost in that type of multiscale geometry.

## 6. CONCLUSIONS

In this work, we showed that, for a microfluidic chamber made from identical arrays of hydrodynamic, passive, traps, the orientation of the flow transporting colloidal particles of different natures have a strong influence, both on the kinetics and the spatial homogeneity of the trapping. In addition, we showed that simple finite element modelling simulations are in excellent agreement with the experiments, and may be used, either to improve the optimization of the trapping for different purposes, or to define some design principles that are sometimes lacking in the literature. Finally, we showed that trapping devices made from single-lithography can be optimized according to their use, while remaining easier to fabricate than more common, two-layer microfluidic devices.

# DECLARATIONS

**Conflict of interest.** The authors declare that they have no conflict of interest.

**Availability of data and materials.** Data and design files are available upon reasonable request.

## Funding

This work has received support from the administrative and technological staff of "Institut Pierre-Gilles de Gennes" (Laboratoire d'excellence : ANR-10-LABX-31, "Investissements d'avenir" : ANR-10-IDEX-0001-02 PSL and Equipement d'excellence : ANR-10-EQPX-34).

JF acknowledges funding from the Agence Nationale de la Recherche (ANR Jeune Chercheur PHAGODROP ANR-15-CE18-0014-01) and from the École Doctorale ED 388 (Chimie Paris Centre) for OM PhD scholarship.

NR acknowledges funding from the École Normale Supérieure de Rennes (ENS Rennes, Contrat Doctoral Spécifique Normalien) for PhD scholarship.

## Authors' contributions.

OM and NR contributed equally to this work. OM and JF designed the experiments; OM performed the experiments. OM, NR, RA, MCJ and JF analyzed the data; NR and RA performed the FEM simulations and analysis, OM, NR, RA, JF, MCJ redacted the manuscript. All authors read and approved the final manuscript.

## Acknowledgements.

We thank the team of engineers of the microfabrication platform of the Institut Pierre-Gilles de Gennes for their technical support along the project. We thank J. Ricouvier (MMN, IPGG, ESPCI) for its support with high-speed camera experiments. We thank T. Podgorsky, T.

SUPPLEMENTARY INFORMATIONS

# Enhancing the capture efficiency and homogeneity of single-layer flow-through trapping microfluidic devices using oblique hydrodynamic streams

O. Mesdjian[1,2‡], N. Ruyssen[3‡], M.-C. Jullien[4], R. Allena[3], J. Fattaccioli[1,2]*

[1] PASTEUR, Département de Chimie, École Normale Supérieure, PSL University, Sorbonne Université, CNRS, 75005 Paris, France

[2] Institut Pierre-Gilles de Gennes pour la Microfluidique, 75005 Paris, France

[3] Institut de Biomécanique Humaine George Charpak, Arts et Métiers ParisTech, Paris, France

[4] Univ. Rennes 1, CNRS, IPR (Institut de Physique de Rennes) UMR 6251, F-35000 Rennes

‡Equal contribution

*Corresponding author e-mail: jacques.fattaccioli@ens.psl.eu

**Hypothesis of laminar flow:**

The Reynold number between traps $R_e$ verifies the relation below:

$$R_e = \frac{\rho \, \Delta x \, V_{moy}}{\mu}$$

Where $V_{moy}$ is the averaged velocity on the cross-surface area of the chamber between two traps. $V_{moy}$ can be approximated by half of the maximal velocity on the velocity profile between traps Figure **Figure S 1B:** $V_{moy} \approx 1 \, mm \, s^{-1}$. Thus:

$$R_e \approx \frac{10^3 . 3 \, 10^{-5} . 10^{-3}}{10^{-3}} = 0{,}03 \ll 1$$

Which confirms the hypothesis of laminar flow in the numerical model.

**Hypothesis of negligible particle-particle interactions:**

We experimentally work with a particle concentration $C = 10^6 \, mL^{-1}$. The minimal volume of the microfluidic chamber (in the case where all the traps are fulfilled) is $Vol_{chamber} = 7{,}3 \, 10^{-12} \, m^3$. Hence, if we consider uniformly distributed particles in the chamber, the number of particles in the chamber is $N_{particles} = 8$. The volume of fluid around an isolated particle is therefore $\frac{Vol_{chamber}}{N_{particles}} = 9{,}1 \, 10^{-11} m^3$. The average distance between two particles $d_{particles}$ verifies:

$$\frac{4}{3} \pi \, d^3_{particles} \approx \frac{Vol_{chamber}}{N_{particles}}$$

Thus:

$$d_{particles} \approx 62 \, \mu m$$

Which is more than 10 times the diameter of our particles. This estimation confirms the hypothesis of negligible particle-particle interactions.

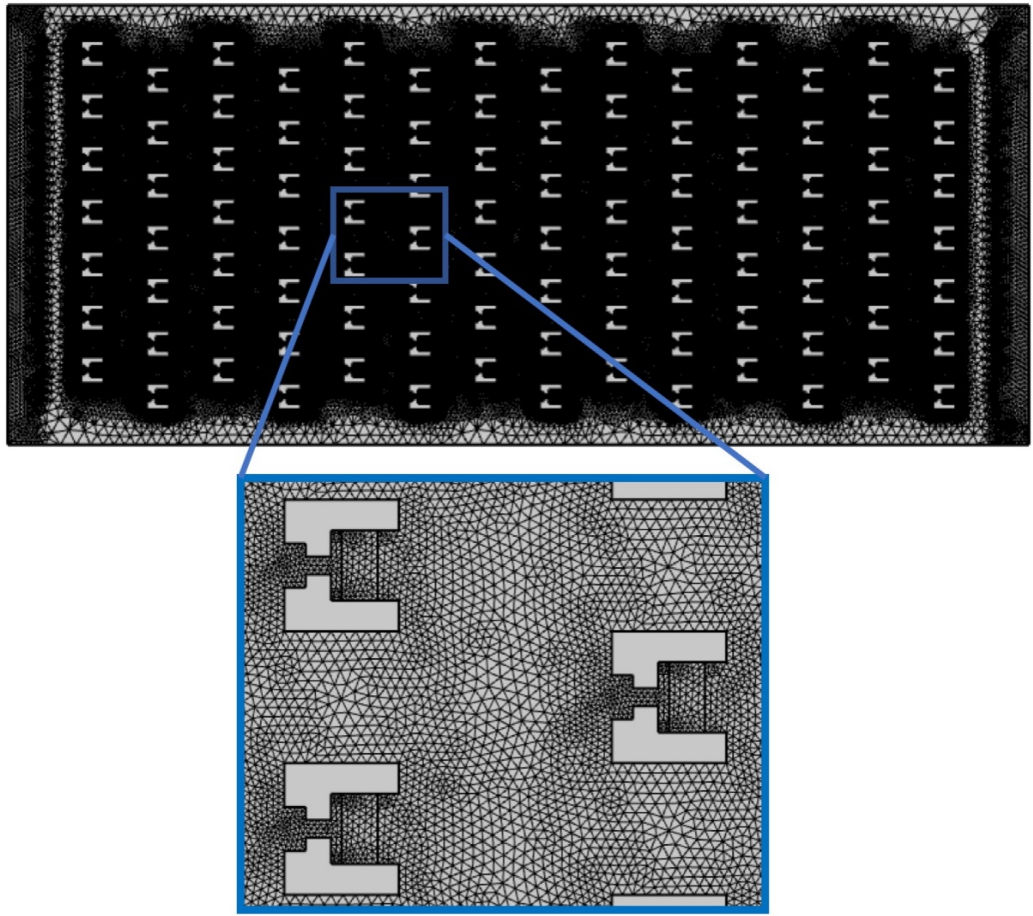

**Figure S 2:** Mesh used for the FEM simulations, composed of linear triangular elements The mesh is refined in the backside opening of the traps.

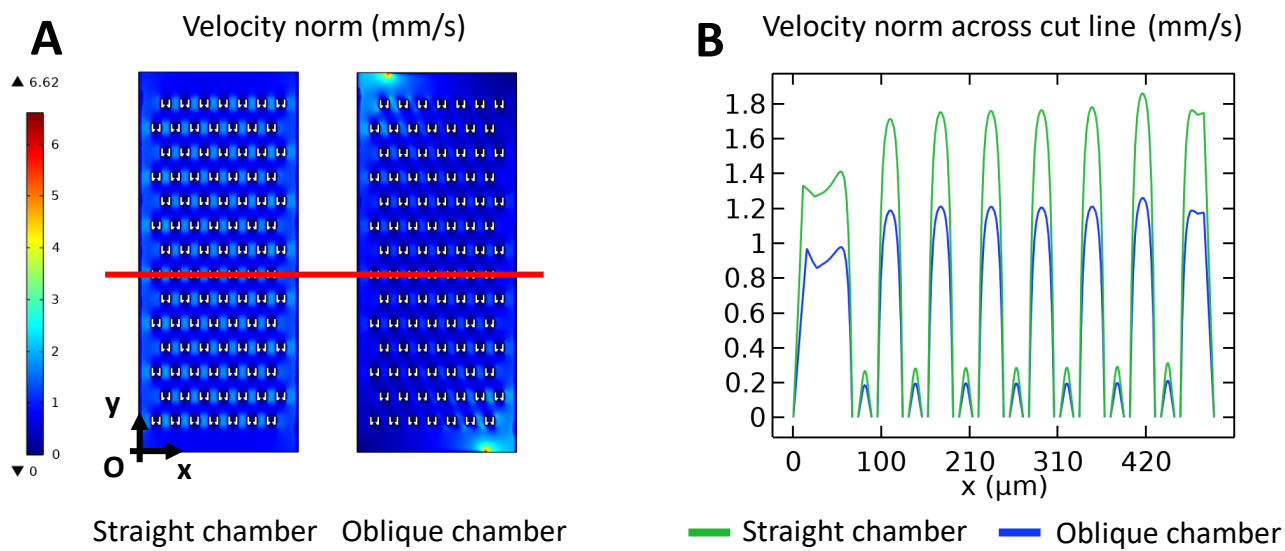

**Figure S 3:** (A) 2D FEM of the hydrodynamic flow within the straight and oblique chamber, for a pressure drop ΔP =100 Pa. To better account for the influence of the top and bottom part of the chip on the flow, a *shallow channel approximation* (h = 14 µm) was used. Colormap corresponds to the hydrodynamic flow velocity expressed in mm.s$^{-1}$. (B) Line profile of the flow velocity (mm.s$^{-1}$) plotted along the red line in (A) for the case of the straight (green) and oblique (blue) chamber. For simulations, traps with a backside opening $w_t$ = 3 µm were used.

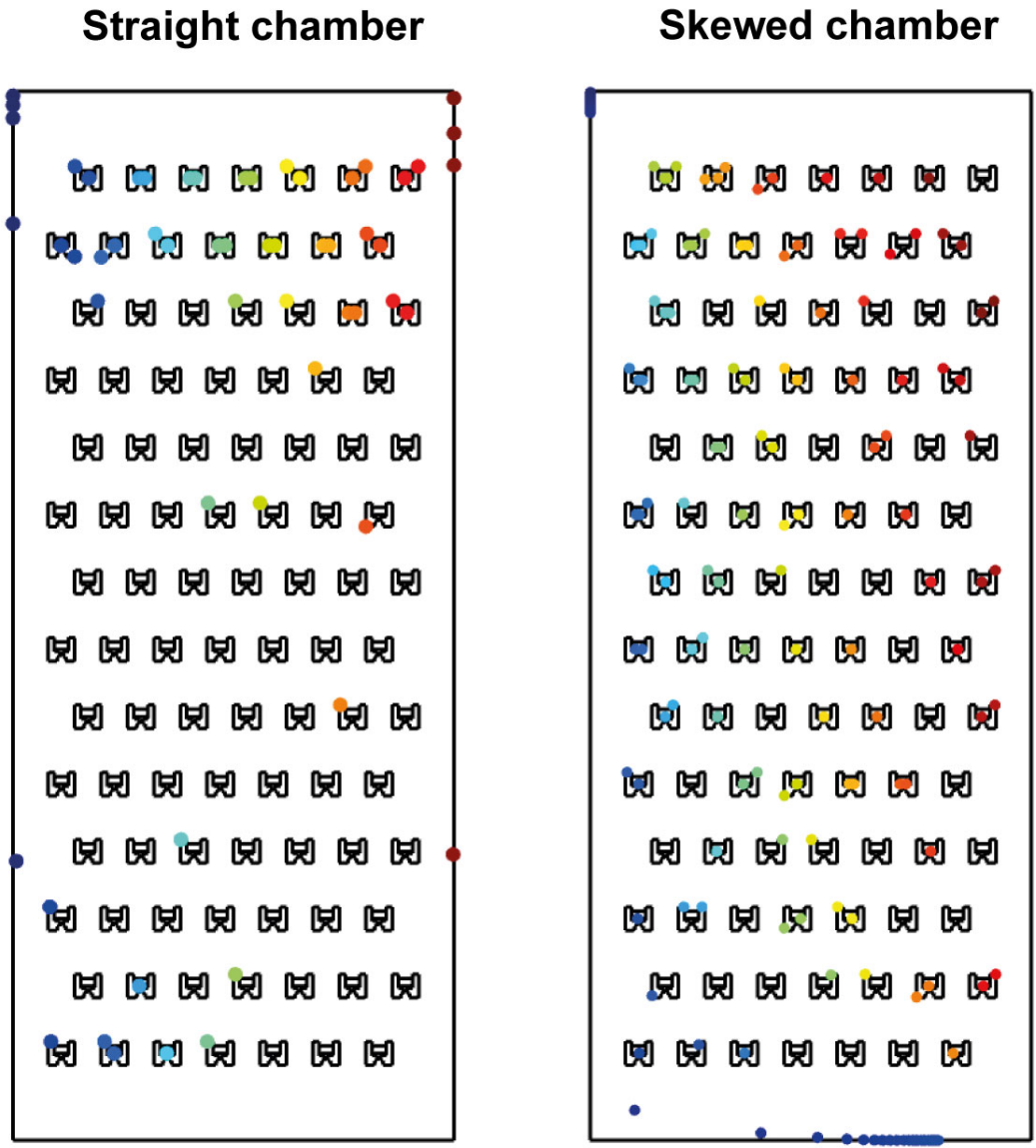

**Figure S 4:** Streamlines stop points and intersections with ($\Gamma_{freeze}$) for the straight (left) and skewed (right) chambers. We notice the rectangular empty region in the middle area of the straight chamber and a more homogeneous filling for the oblique chamber.

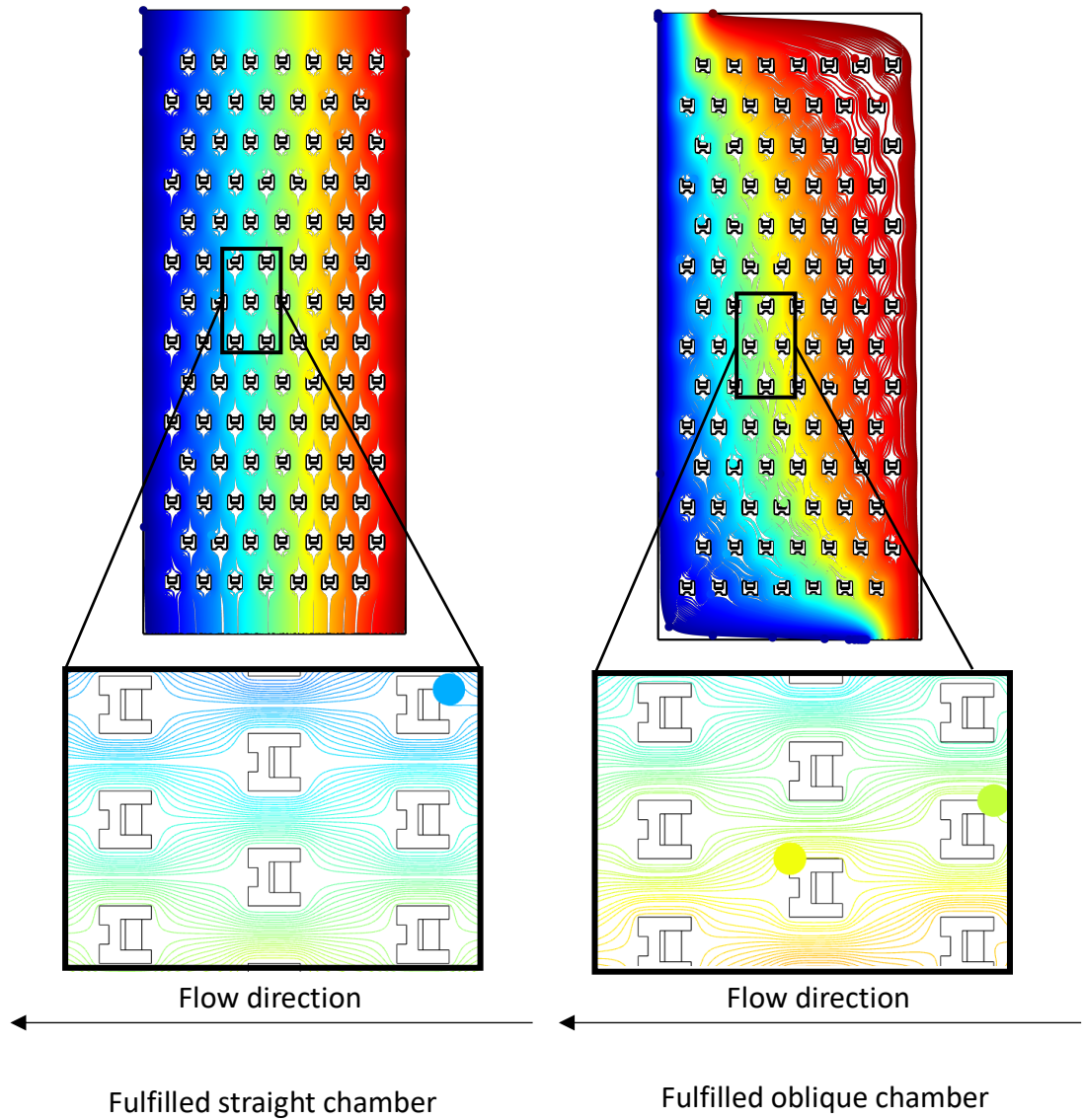

**Figure S 5:** Streamlines and their intersection points with the chamber's walls for the straight (left) and oblique (right) fulfilled chambers. We notice that the streamlines shape is very similar to the streamlines when traps are empty. Streamlines are not physically interrupted by traps but when they cross a trap, the position of their intersection points with ($\Gamma_{freeze}$) is marked by a disc of the same diameter as the experimentally used particles. To facilitate readability, the portions of streamlines crossing traps after the trapping site are not displayed

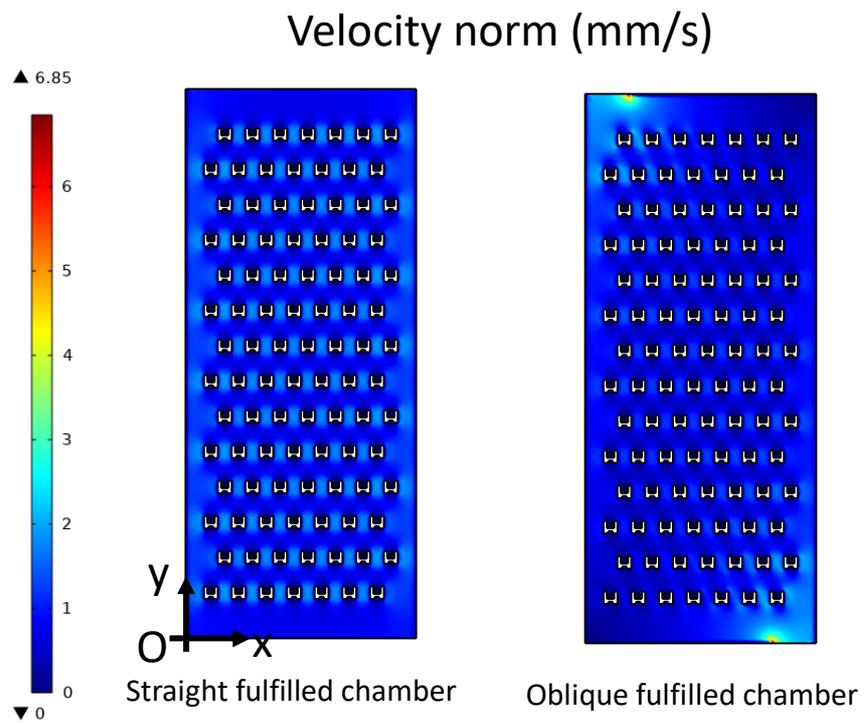

**Figure S 6:** Velocity norm of the fluid flow in the case of fulfilled traps. We notice that this quantity is poorly modified by trapping.

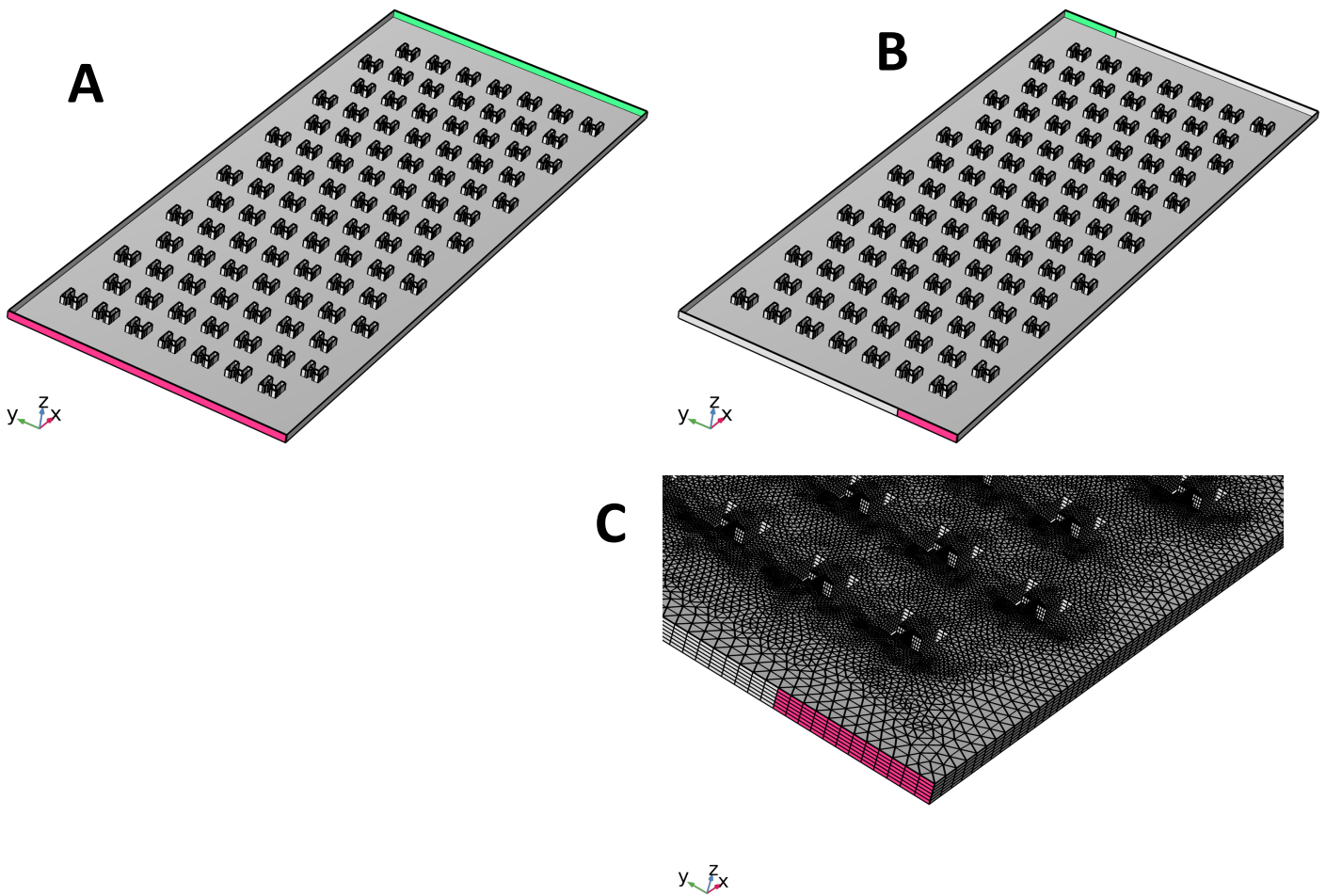

**Figure S 7:** (A) geometry of the 3D model for the symmetric chamber, the inlet surface is in green, and the outlet surface is in purple. (B) geometry of the 3D model for the oblique chamber, the inlet is in green, and the outlet is in purple. (C) the 3D prismatic finite element mesh of the fluid domain. One surface triangular mesh is generated and repeated at different z values. These surface meshes are then vertically connected together to build prismatic 3D volume mesh.

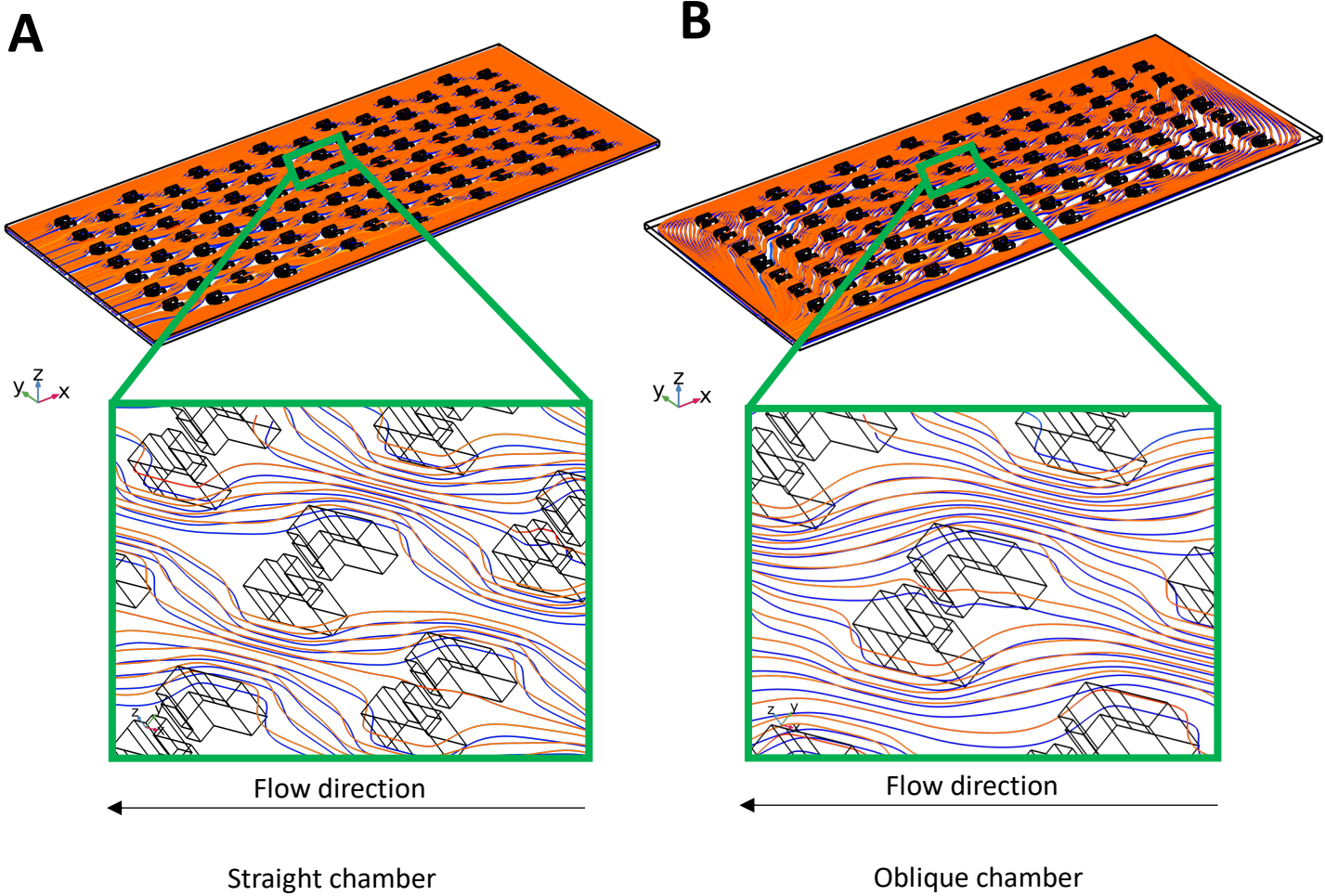

**Figure S 8:** (A) streamlines in the 3D model for the straight chamber and a zoomed view of the streamlines around an empty trap. The streamlines stay by pair together. (B) streamlines in the 3D model for the straight chamber and a zoomed view of the streamlines around a filled trap. The streamlines stay by pair together.

**A1**

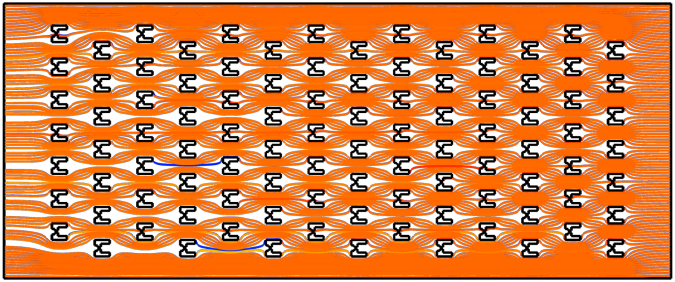

**B1**

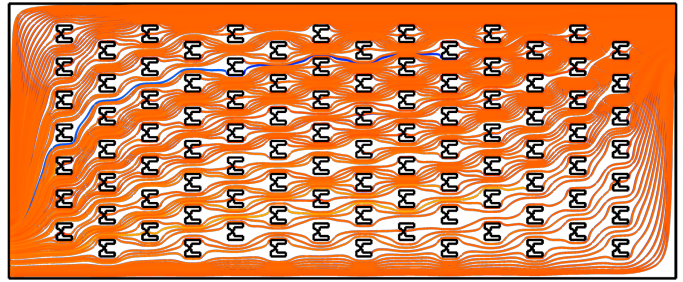

**A2**

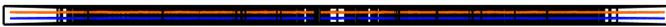

**B2**

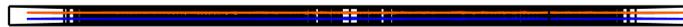

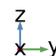

**Figure S 9**: (A) streamlines in the vertical (A1) and horizontal (A2) planar views for the straight chamber of the 3D model. (B) streamlines in the vertical (B1) and horizontal (B2) planar views for the straight chamber of the 3D model.